\documentclass{ws-ijmpe}
\usepackage[super,compress]{cite}
\usepackage{multirow}
\usepackage{float}
\usepackage{tensor}
\usepackage[colorlinks=true,citecolor=blue,urlcolor=blue,linkcolor=red]{hyperref}

\begin{document}

\catchline{}{}{}{}{}
%%%%%%%%%%%%%%%%%%%%%%%%%%%%%%%%%%%%%%%%%%%%
\title{Correlation between the curvature and some properties of the neutron star}

\author{
S. K. Biswal$^{1}$\footnote{subratphy@gmail.com},
H. C. Das$^{2}$,
Ankit Kumar$^{3,4}$,
Bharat Kumar$^{5}$,
Rashmita Jena$^{1,6}$,
Padmalaya Dash$^{7}$,
S. K. Patra$^{7}$}

\address{$^{1}$ Department of Physics, K. K. S Women's  College, Balasore 756001, India}
\address{$^{2}$ INFN Sezione di Catania, Dipartimento di Fisica, Via S. Sofia 64, 95123 Catania, Italy}
\address{$^{3}$ Institute of Physics, Sachivalaya Marg, Bhubaneswar 751005, India
}
\address{$^{4}$ Homi Bhabha National Institute, Training School Complex, Anushakti Nagar, Mumbai 400094, India}
\address{$^{5}$ Department of Physics $\&$ Astronomy, National Institute of Technology, Rourkela 769008, India }
\address{$^{6}$ Department of Physics, Fakir Mohan University, Balasore, Odisha, 756019, India}
\address{$^{7}$ Department of Physics, Siksha O Anusandhan Deemed to be University, Bhubaneswar 751030, India}

\maketitle
\abstract{
According to the general theory of relativity, a massive body induces curvature in the surrounding spacetime. In this study, the surface curvature (SC) of neutron stars is computed using various curvature quantities derived from the relativistic mean-field (RMF), density-dependent RMF (DDRMF), and Skyrme-Hartree-Fock equations of states. Neutron star properties, including mass, radius, compactness, and central density, are calculated utilizing the Tolman-Oppenheimer-Volkoff equations. The analysis reveals a significant cubic correlation between the SC and compactness for the canonical 1.4 $M_{\odot}$ neutron star, with a correlation coefficient of 0.99, indicating an almost linear relationship. A similarly significant inverse cubic correlation is observed between the SC and the radius of the canonical star. However, these correlations diminish for the maximum mass NS. Furthermore, a universal relation between the SC and the dimensionless tidal deformability ($\Lambda$) for the canonical neutron star is established. Using the tidal deformability constraint of GW170817 ($\Lambda_{1.4} = 190_{-120}^{+390}$), the surface curvature is limited to SC$_{1.4} (10^{14}) = 2.87^{+0.30}_{-0.78}$ at a confidence level 90\%. Furthermore, the tidal deformability constraint of the secondary component in the GW190814 event ($\Lambda_{1.4} = 616_{-158}^{+273}$) \cite{RAbbott_2020} offers a more stringent limit, with the result of SC$_{1.4} (10^{14}) = 2.03^{+0.27}_{-0.36}$. These findings indicate that the GW190814 event imposes more rigorous constraints on SC compared to GW170817. 

}
%%%%%%%%%%%%%
%%%%%%%%%%%%%
%\begin{document}
%\maketitle
\flushbottom
%%%%%%%%%%%
%%%%%%%%%%%%%%%%%%%%%%
\section{Introduction}
\label{intro}
%%%%%%%%%%%%%%%%%%%%%%
Investigation of the fundamental interactions at supra-saturation density is presently regarded as a significant topic in fundamental physics. Terrestrial experiments have achieved the creation of dense matter up to several times the saturation density \cite{Danielewicz_2014, Steiner_2010}; however, this remains insufficient to fully understand nuclear interactions at high density and isospin asymmetry. Consequently, this experimental limitation necessitates the use of neutron stars (NS) as a unique probe to study nuclear interactions in high-density regimes \cite{latt04}. Neutron stars, prominent members of the compact object category within the observable universe, possess central densities that range from 5 to 6 times that of matter found at the nuclear center \cite{baym79,latt04}. Terrestrial experiments and analyses of finite nuclear data predominantly provide insights into the equation of state (EOS)  around saturation density, while higher-density regions remain largely unexplored. Theoretical models predict a wide range of behaviors for the EOS in these unexplored regions, resulting in markedly different predictions for the global properties of neutron stars. These include maximum mass, radius, tidal deformability, and moment of inertia, which serve as crucial parameters to impose stringent constraints on the nature of the EOS \cite{latt04}.

In recent years, the tidal deformability of the first binary NS merger event, GW170817 \cite{abbo18,abbo17}, has provided insight into the nature of the material enclosed within an NS. The precise composition of the NSs remains unclear. Conventional models posit that NSs primarily consist of neutrons, with a mixture of protons and electrons present to maintain $\beta$-equilibrium and charge neutrality. The extremely high density inside NSs facilitates various exotic phenomena, such as hyperon production \cite{amba60,glen85,glen91}, kaon condensation \cite{glen85,glen98,glen99,chri00,neha12,neha13}, quark deconfinement \cite{coll75,xia16, Xiab16, Xia17,dexh18}, and the presence of dark matter particles \cite{Panotopoulos_2017, Das_2019, Li_2012, Das_2020, Ellis_2018, Bhat_2019, Ciarcelluti_2011, Leung_2011}. Recent data from the Neutron Star Interior Composition Explorer (NICER) \cite{Raaijmakers_2019, Miller_2019, Bogdanov_2019, Bilous_2019, Riley_2019, Guillot_2019} have imposed stringent constraints on the mass and radius of NSs. This simultaneous measurement of NS mass and radius can constrain the nuclear EOS.

In the literature, various approaches have defined the correlation between tidal deformability and the radius of the canonical star (a star of mass $1.4 \ M_\odot$) \cite{Malik_2018, Hebler_2010, Lim_2018, Most_2018, Fattoyev_2018, Tews_2018, Annala_2018, Zhang_2018, Dietrich_2020, Steiner_2010, Capano_2019}, and this issue is now settled. However, the value of the correlation coefficient remains a matter of debate; different theoretical frameworks yield varying ranges of correlation coefficients. Furthermore, a correlation has been observed between the skin thickness of the heavy neutron-rich nuclei ($^{208}$ Pb) and various NS properties \cite{Fatttoyev_2012, Horowitz_2001}. This is attributable to the fact that the slope of the symmetry energy determines the pressure of the neutron-rich skin and the radius of the NS. 
Li et al. showed that the connection between $K_{\rm sym}$ and slope $L$ constrains the high-density behavior of symmetry energy $E_{\rm sym}$ \cite{Baoli_2020}. Alam et al. found a strong link between NS radii and the slope of nuclear matter incompressibility \cite{Alam_2016} using expression $M_0 + \beta L_0$, aiding radius constraints when $M_0$ and $L_0$ are known. Wei et al. \cite{Wei_2020} correlated NS radii, tidal deformability, and beta-stable matter pressure within Brueckner-Hartree-Fock formalism. These findings help indirectly constrain the NS EOS and offer insights into high-density nuclear matter. Inspired by these, our research explores a new link between NS surface curvature and properties like radius, compactness, and tidal deformability. Surface curvature is crucial for its connection to gravitational potential and spacetime curvature, key elements of strong-field gravity. For example, Rosi {\it et al.} \cite{Rosi15} discusses atom interferometry for measuring curvature and gravitational factors applicable to NSs. Curvature is a crucial quantity in GR, defines gravitational field strengths, and can aid EOS constraint and deepen gravitational theory comprehension in strong-field contexts. Our study establishes a universal link between NS surface curvature and features like compactness and tidal deformability, utilizing observational data like GW170817. These relations indirectly constrain curvature and inform on neutron star maximum mass. By using various EOS models (relativistic and non-relativistic), we reduce model bias and enhance understanding of NS properties. Although the universal correlation doesn't directly constrain the EOS, it serves as an auxiliary tool for pinpointing EOS models aligned with observations and gravitational principles.

According to general relativity, the energy and momentum of any matter or radiation present in the universe are intricately linked to the curvature of space-time. The degree of this curvature is contingent upon the distortion caused by a massive object within space-time, analogously to the deformation observed in a trampoline. To elucidate this unified theory of gravitation through geometric algebra, specific mathematical quantities such as compactness ($C\equiv M/R$), the Riemann tensor, the Ricci scalar, and the Kretschmann scalar have been defined. A more comprehensive explanation and derivation of these mathematical quantities, which encapsulate extensive information regarding curvature, can be found in Ref. \cite{Kazim_2014}. Among these quantities, those referenced in $\cal{K}$ and $\cal{W}$ are particularly significant in quantifying space-time curvature both within and outside stellar objects.

Modern observational instruments enable deeper exploration into the strong-field regime of neutron stars. The core is approximately fifteen times denser than the surface, indicating that the majority of the matter is concentrated in the core, thereby complicating the precise determination of the NS mass-radius profile. However, the star's compactness and surface curvature increase radially towards the surface. This possibly explains why the maximum mass-radius measurement is more prominent for the equation of state than for gravity. This work presents recent methodologies for studying NS curvature. The unconstrained gravity of the NS within the general relativity (GR) framework is quantified in Ref. \cite{Kazim_2014}. They estimated NS curvature and revealed that GR is less extensively tested across the star compared to EOS. Additionally, it was found that the radial variation of the Weyl tensor follows a power law throughout the star. Xiao et al. \cite{Xiao_2015} utilized both relativistic mean-field and Skyrme-Hartree-Fock EOSs to determine that symmetry energy significantly affects the curvature of lighter NSs but has minimal impact on more massive NSs. Moreover, to quantify deviations from GR in the strong-field regime, a comprehensive understanding of the matter within the NS is essential. In our previous analysis, we quantified the curvatures of NS with and without the inclusion of DM \cite{Das_2021}. Our findings demonstrated that DM exerts a notable influence on the curvatures and compactness of NS. Therefore, the objective of this current study is to delve deeper into this curvature and establish correlations with specific NS properties. Such correlations between curvatures and various bulk properties of the NS are essential for a comprehensive understanding of their nature.

%%%%%%%%%%%%%%%%%%%
\section{Framework}
\label{method}
%%%%%%%%%%%%%%%%%%%
This section delineates various quantities essential for calculating the curvature in both the interior and exterior regions of neutron stars across different equations of states. To examine the properties of NS, we employ relativistic mean-field (RMF), Skyrme-Hartree-Fock (SHF), and density-dependent RMF (DDRMF) EOSs. Over the last few decades, the aforementioned models have established a robust foundation for predicting properties of both finite and infinite nuclear matter. Unlike non-relativistic models, the RMF model adheres to the causal limit even at very high densities \cite{dutr14}. The causal nature of the RMF model facilitates a seamless transition from finite nuclei to NS, which are characterized by extremely high-density environments. Furthermore, the SHF models considered here do not violate the causality limit for masses below $2M_\odot$.

%%%%%%%%%%%%%%%%%%%%%%%%%%%%%%%%%%%%%%%%%%%%%%%%%%%%%%
\subsection{Mathematical form for different curvatures}
\label{fcurvature}
%%%%%%%%%%%%%%%%%%%%%%%%%%%%%%%%%%%%%%%%%%%%%%%%%%%%%%
In this section, we use curvature quantities from Ref. \cite{Kazim_2014, Das_2021}: the Ricci scalar, full contraction of the Ricci tensor, Kretschmann scalar, and the full contraction of the Weyl tensor to measure space-time curvature. The Ricci tensor, Weyl tensor, and Kretschmann scalar analyze curvature in a Riemannian manifold. The Ricci tensor indicates volume change along a geodesic path, while its trace gives the Ricci scalar. The Weyl tensor, the traceless part of the Riemann tensor, describes shape deformation due to tidal effects. In general relativity, the Ricci tensor is zero in a vacuum; the Weyl tensor is not. The Kretschmann scalar, related to the Riemann tensor, identifies singularities with infinite space-time curvature. These space-time curvature tensors are defined below.\\

%{\color{blue} The Ricci tensor, the Weyl tensor, and the Kretschmann scalar are used to analyze the space-time curvature in a Riemannian manifold. The Ricci tensor measures the change in volume encountered by a body moving along a geodesic path. The contraction of the Riemann tensor gives the Ricci tensor, and the curvature scalar or Ricci scalar is defined as the trace of the Ricci tensor. The Weyl tensor represents the traceless component of the Riemann tensor. The Weyl tensor describes the deformation in shape due to the tidal effect. In the general relativity context, the Ricci tensor vanishes in vacuum or free space, whereas the Weyl tensor is non-zero. The Kretschmann scalar has properties similar to those of the Riemann tensor, and the square of the Kretschmann scalar represents the full contraction of the Riemann tensor. In general relativity, the Kretschmann scalar is used to identify singurality -- the region of infinite space-time curvature. These space-time curvature tensors are defined below.}

The Ricci scalar
\begin{equation}
{\cal R}(r)=8\pi\bigg[{\cal{E}}(r) -3 P(r)\bigg],
\label{RS}
\end{equation}
the full contraction of the Ricci tensor
\begin{equation}
{\cal J}(r) \equiv \sqrt{{\cal R}_{\mu \nu} {\cal R}^{\mu \nu}} = 8\pi \left[ {\cal{E}}^2(r) + 3P^2(r)\right]^{1/2},
\label{RT}
\end{equation}
the Kretschmann scalar 
\begin{eqnarray}
{\cal{K}}(r)&\equiv&\sqrt{{\cal{R}}^{\mu\nu\rho\sigma}{\cal{R}}_{\mu\nu\rho\sigma}}
\nonumber\\
&=& \bigg[(8\pi)^2\left \{3{\cal{E}}^2(r)+3^2(r)
+2P(r){\cal{E}}(r)\right\}
-\frac{128{\cal{E}}(r)m(r)}{r^3} +\frac{48m^2(r)}{r^6}\bigg]^{1/2},
\label{KS}
\end{eqnarray}
and the full contraction of the Weyl tensor
\begin{equation}
{\cal W}(r) \equiv \sqrt{{\cal C}^{\mu \nu \rho \sigma }{\cal C}_{\mu \nu \rho \sigma}} = \bigg[\frac43 \left( \frac{6m(r)}{r^3} - 8\pi {\cal{E}}(r) \right)^2\bigg]^{1/2},
\label{WT}
\end{equation}
%%%%%%%%%%%%%%
where ${\cal E}$ and $P$ are the energy density and pressure of the systems. The $m(r)$ is the enclosing mass with areal radius $r$. To obtain the mass and radius of the star, one needs to solve the following Tolman-Oppenheimer-Volkoff equations ~\cite{TOV1, TOV2}
%%%%%%%%%%%%%%%%
\begin{eqnarray}
\frac{dm(r)}{dr}&=&4\pi r^2 {\cal E}(r),
\nonumber\\
\frac{dP(r)}{dr}&=&-\Big[P(r)+{\cal E}(r)\Big] \frac{d\Phi}{dr},
\nonumber \\
\frac{d\Phi (r)}{dr}&=&\frac{m(r)+4\pi r^3P(r)}{r[r-2m(r)]}.
\end{eqnarray}
%%%%%%%%%%%%%%
The coupled differential equations can be solved using the boundary conditions $r=0$, $P=P_c$, and $r=R$, $P=0$.

\begin{figure*}
\centering
\includegraphics[width=0.52\textwidth]{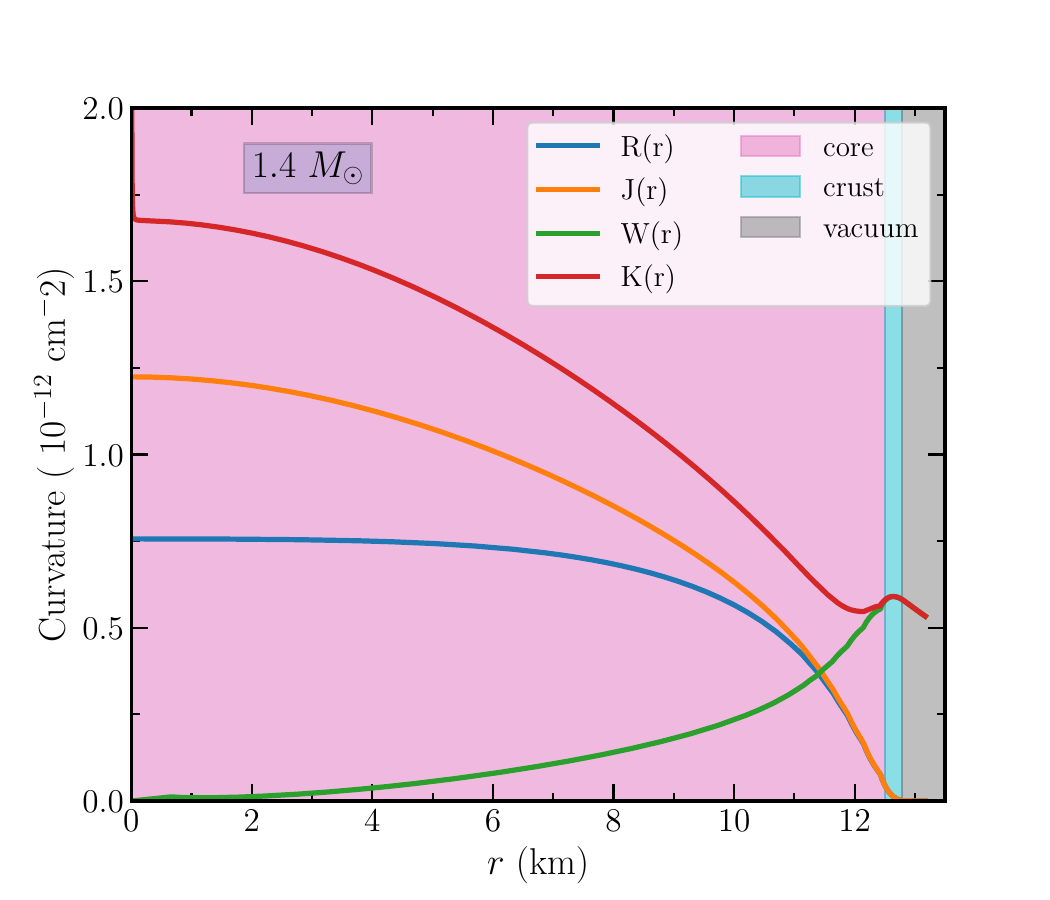}%%
\includegraphics[width=0.52\textwidth]{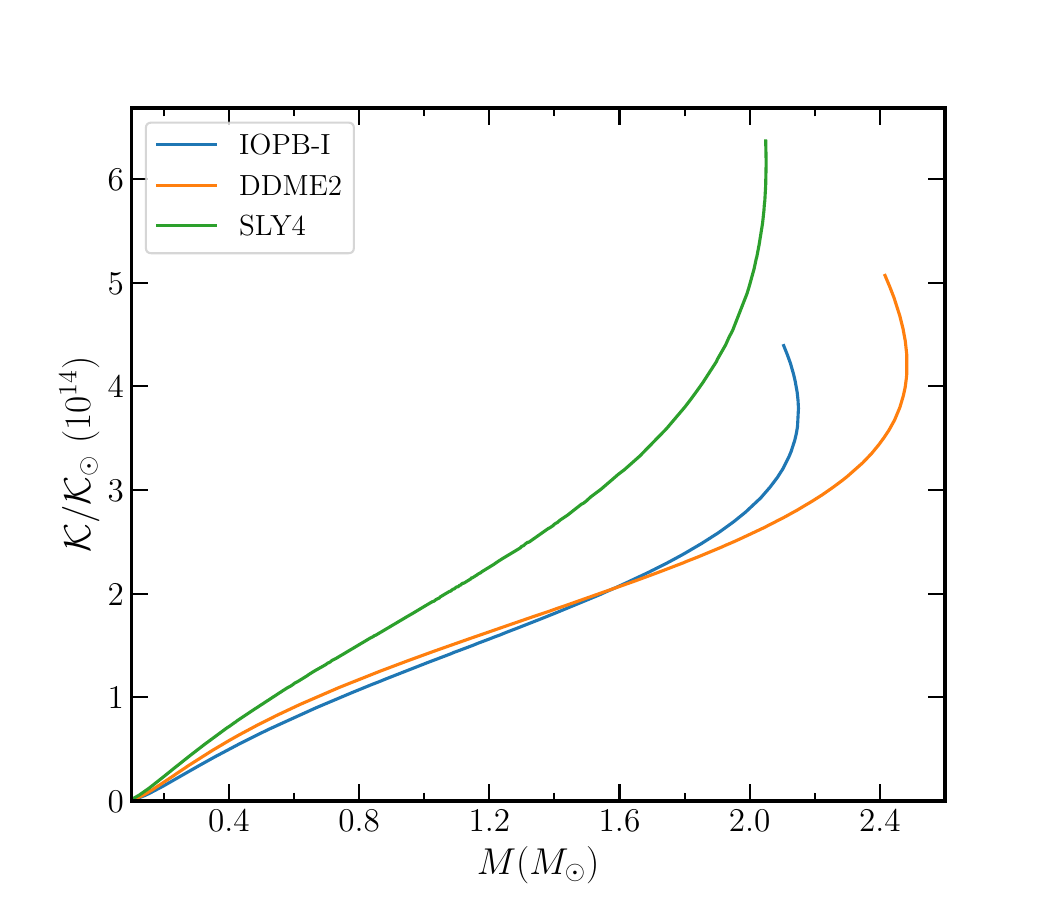}
\caption{{\it Left:} The curvature quantities within canonical star are shown for IOPB-I EOS, reproduced from Ref. \cite{Kazim_2014}. {\it Right:} Surface curvatures (${\cal{K}}(R)/{\cal{K}}_\odot$) for IOPB-I, DDME2, and SLY4 EOSs.}
\label{fig_curv}
\end{figure*}

At the surface $m\rightarrow M$ due to $r \rightarrow R$. The Ricci tensor and Ricci scalar vanish outside the star because they depend on the ${\cal{E}}(r)$, $P(r)$, which are zero outside the star. However, there exists a non-zero component of the Riemann tensor; $\tensor{{\cal R}}{^1_{010}}=-\frac{2M}{R^3}=- \xi$, even external to the star \cite{Kazim_2014, Xiao_2015}. Consequently, the Riemann tensor is a more pertinent quantity for assessing the curvature of the stars. The Kretschmann scalar is defined as the square root of the full contraction of the Riemann tensor. The vacuum value for both $\cal{K}$ and $\cal{W}$ is $\frac{4\sqrt{3}M}{R^3}$, which is evident from Eqs. (\ref{KS}) and (\ref{WT}). Therefore, $\cal K$ and $\cal W$ can be considered reasonable measures for quantifying the curvature within the star. The SC is defined as the ratio of curvature at the surface of the NS ${\cal{K}}(R)$ to that of the Sun ${\cal{K}}_\odot$, denoted as SC $={\cal{K}}(R)/{\cal{K}}_\odot$. This ratio ${\cal{K}}(R)/{\cal{K}}_\odot\approx10^{14}$; in essence, the NS curvature is $10^{14}$ times greater than that of the Sun (refer to Fig. \ref{fig_curv}).

%%%%%%%%%%%%%%%
%%%%%%%%%%%%%

Figure \ref{fig_curv} illustrates the curvature quantities as functions of radius for a canonical neutron star. In the left panel, the various curvature quantities—${\cal{R}}(r)$, ${\cal{J}}(r)$, ${\cal{W}}(r)$, and ${\cal{K}}(r)$—are plotted for the IOPB-I EOS. It is observable that all curvature quantities tend to converge towards comparable values at the crust-core boundary; however, their magnitudes and trends vary across the core, crust, and vacuum regions. At the center of the star ($r = 0$), ${\cal{K}}(r)$ displays the highest magnitude among the curvature measures, whereas $\cal{W}$ presents a contrary trend as compared to the other curvature quantities. The surface curvature ${\cal{K}}(R)$ increases progressively from the center towards the star's surface, mirroring the intensification of the gravitational field as one moves outward. The right panel of Fig. \ref{fig_curv} portrays the surface curvatures ${\cal{K}}(R)/\cal{K}_{\odot}$ for three different EOSs—IOPB-I, DDME2, and SLY4—plotted against neutron star mass. For a neutron star with a fixed mass (e.g., $2 M_{\odot}$) and central density, the surface curvature remains invariant. Conversely, for stars with varying central densities, surface curvature amplifies with increasing mass. Among the EOSs, the softer EOS, SLY4, forecasts a higher magnitude of surface curvature compared to the stiffer EOSs like IOPB-I. This indicates that softer EOSs result in more compact stars with stronger gravitational fields at the surface, leading to greater curvatures. The analysis reveals that gravitational curvature in neutron stars can be up to $10^{14}$ times more intense than that around the Sun. Moreover, ${\cal{K}}(r)$ escalates with the increasing density towards the center of the neutron star. The value of ${\cal{W}}(r)$ originates from zero at the center ($r = 0$) and incrementally rises radially outwards towards the crust, obeying a power law behavior (${\cal{W}}(r) \propto r^2$). This suggests that while ${\cal{K}}(r)$ reaches its peak in the denser core regions, ${\cal{W}}(r)$ gains significance closer to the star's surface. The behavior of ${\cal{K}}(r)$ near the core is predominantly attributed to the extreme gravitational fields that elude weak-field approximations, as noted in Ref. \cite{Psaltis_2008}. Observations from our solar system, which predominantly probe weak-field gravity, fail to adequately constrain the strong-field regime within neutron stars. The curvature magnitude generated by a neutron star's gravitational field is on the order of $10^{14}$ compared to that of the Sun, indicating the extraordinary intensity of the gravitational field within a neutron star. This intensity constrains our ability to govern the curvature in the inner regions of the star. Conversely, ${\cal{W}}(r)$ attains its maximum value near the crust, where the implications of general relativity (GR) are significant but manageable. This underscores the potential limitations of GR under extreme strong-field conditions while affirming its efficacy in weaker fields near the star's surface. These studies highlight that GR is less rigorously examined in the extreme environments of neutron stars as compared to the constraints available for the equations of state.
%%%%%%%%%%%%%%%%%%%%%%%%%%%%%%%%%
\section{Results and Discussions} 
\label{result}
%%%%%%%%%%%%%%%%%%%%%%%%%%%%%%%%%
%%%%%%%%%%%%%%%%%%%%%%%%%%%%%%%%%%%%%%%%
\subsection{Equation of states of the NSs}
%%%%%%%%%%%%%%%%%%%%%%%%%%%%%%%%%%%%%%%%
The structure of an NS is divided into four regions: the outer crust, inner crust, outer core, and inner core. The outer crust comprises nuclei arranged in a body-centered-cubic lattice immersed in a free electron gas. From the outer to the inner crust, the density intensifies, causing the nuclei to become increasingly neutron-rich to the extent that neutrons start to drip from them. Consequently, the inner crust encompasses free electron and neutron gases and manifests various pasta structures. Utilizing nuclear models, the neutron drip density demarcates the boundary between the outer and inner crust. However, the transition density from the crust to the core remains notably uncertain and heavily model-dependent. When the mean density attains approximately half of the NM saturation density, the lattice structure dissolves due to energetic constraints, and the system transitions to a liquid phase. To calculate the nuclear interaction of this region mainly for the core part, we have several formalisms (for relativistic to non-relativistic system) that have already been proposed \cite{Kumar_2018}. The details of those models, such as RMF, SHF, and DD-RMF, are described briefly in the following.

RMF is an effective field theory motivated model in which nucleons interact with each other through the exchange of various effective mesons like $\sigma$, $\omega$, and $\rho$. The effective Lagrangian is given by
\cite{rein86,mill72,furn87,ring96}\\
\begin{eqnarray}
&&{\cal L}=\sum_i\overline{\psi}_i\bigg(
i\gamma^{\mu}\partial_{\mu}-m+g_{\sigma}\sigma -g_{\omega}\gamma_\mu
 \omega^ \mu -\frac{1}{2}g_{\rho}\gamma \bigg)
\psi_i \nonumber \\
&&+ \frac{1}{2}\partial_{\mu}\sigma\partial^{\mu}\sigma -m_{\sigma}^2\sigma^2
\left(\frac{1}{2}+\frac{\kappa_3}{3!}\frac{g_{\sigma}\sigma}{m}
+\frac{\kappa_4}{4!}\frac{g_{\sigma}^2\sigma^2}{m^2}\right)
 \nonumber \\
&&  - \frac{1}{4}\omega_{\mu\nu}\omega^{\mu\nu} +\frac{1}{2}m_{\omega}^2
\omega_{\mu}\omega^{\mu}\left(1+\eta_1\frac{g_{\sigma}\sigma}{m}
+\frac{\eta_2}{2}\frac{g_{\sigma}^2\sigma^2}{m^2}\right) +\frac{1}{4!}\zeta_0 \left(g_{\omega}\omega_{\mu}\omega^{\mu}\right)^2
\nonumber \\
&&-\frac{1}{4}R_{\mu\nu}R^{\mu\nu} +\frac{1}{2}m_{\rho}^2
R_{\mu}R^{\mu}\left(1+\eta_{\rho}
\frac{g_{\sigma}\sigma}{m} \right)
+\Lambda_v R_{\mu}R^{\mu}
(\omega_{\mu}\omega^{\mu}) \nonumber\\
&&+\sum_l\overline{\psi}_l\left(
i\gamma^{\mu}\partial_{\mu}-m_l\right)\psi_l,
\end{eqnarray}
where i use for nucleons ($n, p$), and $l$ represents the leptons. $\omega_{\mu\nu}$ and $R_{\mu\nu}$ are defined as
$\omega_{\mu\nu}=\partial_\mu \omega_\nu-\partial_\nu \omega_\mu$ and
$R_{\mu\nu}=\partial_\mu R_\nu-\partial_\nu R_\mu$ respectively. The $\omega_{\mu\nu}$ and $R_{\mu\nu}$ are the field tensors of the $\omega$--meson and $\rho$--meson respectively. The strength of the interaction is decided by the coupling constants $g_{\sigma}$, $g_{\omega}$, $g_{\rho}$, $\Lambda_v$. Every parameter set of the RMF model has a unique set of coupling constant values. So, the interaction is uniquely defined in each parameter set. From the Lagrangian, we can calculate the energy density and pressure from the energy-momentum tensor \cite{glenb97}.

The SHF model is the non-relativistic version of the RMF model, where nucleons are treated as non-relativistic particles. The interaction of the nucleons is governed by the non-relativistic Schrodinger equation. The SHF model starts with effective Hamiltonian H is defined as
\begin{eqnarray}
&&    H = \frac{\hbar}{2m}\tau_N + \frac{1}{4}t_0[(2+x_0)\rho^2_N-(2x_0+1)(\rho^2_p+\rho^2_n)]\nonumber\\
&&    +\frac{1}{24}t_3\rho^{\alpha}[(2+x_3)\rho^2_N-(2x_3+1)(\rho^2_p+\rho^2_n)]\nonumber\\
&&    +\frac{1}{8}\tau_N \rho_N[t_1(2+x_1)+t_2(2+x_2)]\nonumber\\
&&    +\frac{1}{8}[t_2(2x_2+1)-t_1(2x_1+1)](\tau_p\rho_p+\tau_N\rho_N)\nonumber\\
&& +\frac{1}{32}[3t_1(2+x_1)-t_2(2+x_2)](\nabla_{\rho N})^2\nonumber\\
&& -\frac{1}{32}[3t_1(2x_1+1)+t2(2x_2+1)][(\nabla_{\rho p})^2+(\nabla_\rho N)^2]\nonumber\\
&& +\frac{1}{2}W_0[J_N.\nabla_{\rho N}+J_p.\nabla_{\rho p}+J_n.\nabla_{\rho n}]\nonumber\\
&& -\frac{1}{16}(t_1x_1+t_2x_2)J^2_N+\frac{1}{16}(t_1-t_2)[J^2_p+J^2_n]\nonumber\\
&&+\frac{1}{2}e^2\rho_p(r)\int \frac{\rho_p(r^{\prime})d^3r^{\prime}}{|r-r^{\prime}|}-\frac{3}{4}e^2\rho_p(r) \left ( \frac{3\rho_p(r)}{\pi}\right)^{1/3},
\end{eqnarray}
where n and p correspond to neutron and proton. The term $\rho_N$, $\tau_N$ and $J_N$ are defined as, $\rho_N = \rho_n+\rho_p$, $\tau_N = \tau_n+\tau_p$ and $J_N = J_n+J_p$.
From the Hamiltonian, we calculated energy density and pressure.

The DDRMF model is based on the same formulation of the RMF model, except the coupling constants are density-dependent. The equations for density-dependent coupling are given by \cite{Banik_2014}
\begin{eqnarray}
    g_{\alpha N} = g_{\alpha N}(n_0) f_{\alpha}(n_b/n_0), \nonumber\\
    f_{\alpha}(n_b/n_0) = a_{\alpha} \frac{1+b_{\alpha}(x+d_{\alpha})^2}{1+c_{\alpha}(x+d_{\alpha})^2},
\end{eqnarray}
where $\alpha = \sigma, \omega$, $n_0$ is the saturation density and $n_b$ is the baryon density. The density-dependent coupling $g_{\rho N}$ for $\rho-$meson is defined as
\begin{eqnarray}
    g_{\rho N} = g_{\rho N}  exp[-a_{\rho} (n_b/n_0 -1)].
\end{eqnarray}
%%%%%%%%%%%%%%%
The procedure for calculating energy density and pressure is similar to the RMF model. The detail about the DDRMF model can be found in the Refs.\cite{Niksic_2002,Lalazissis_2005,Banik_2014,Fortin_2016,Wei_2020}.

In the present calculation, we used three different category with 30 parameter sets of the RMF models, such as HS \cite{horo81}, LA \cite{furn87}, H1 \cite{hadd99}, LZ \cite{rein89}, NL3 \cite{lala97}, NL3* \cite{lala09}, NL-SH \cite{sharma_1993}, NL1 \cite{rein86}, GM1 \cite{glen91}, GL85 \cite{glen85}, GL97 \cite{glenb97}, NL3-II \cite{lala97}, NLD \cite{furn87}, NL-RA1 \cite{rash01}, TM1 \cite{suga94}, TM2 \cite{suga94}, PK1 \cite{long04}, FSU \cite{todd05}, FSU2 \cite{weic14}, IUFSU \cite{Fattoyev_2010}, IUFSU* \cite{fatt10}, SINPA \cite{mond16}, SINPB \cite{mond16}, G1 \cite{frun97}, G2 \cite{frun97}, G3 \cite{bhar17}, IOPB-I \cite{Kumar_2018}, FSUGarnet \cite{Chen_2015}, FSU2R \cite{Tolos_2016}, and FSU2H \cite{Tolos_2016}; 15 SHF models such as BSk20, BSk21 \cite{Goriely_2010}, BSk22, BSk23, BSk24, BSk25 \cite{Goriely_2013}, KDE0v1 \cite{Agrawal_2005}, SkI2, SkI3, SkI5 \cite{Reinhard_1995}, SkI6 \cite{Nazarewicz_1996}, SLY2 \cite{Chabanat_1995}, SLY4 \cite{Chabanat_1998}, SL230a \cite{Chabanat_1997}, and Rs \cite{Friedrich_1986}; and 6 DDRMF parameter sets such as DD2 \cite{Banik_2014}, DD2Y \cite{Fortin_2016}, DDME1 \cite{Niksic_2002}, DDME2 \cite{Lalazissis_2005}, DDME2Y \cite{Fortin_2016}, and DDLZ1 \cite{Wei_2020} are used to investigate various properties of NSs and to explore the existing correlations between different quantities.
%%%%%%%%%%%%%
\begin{figure*}
\centering
\includegraphics[width=0.52\textwidth]{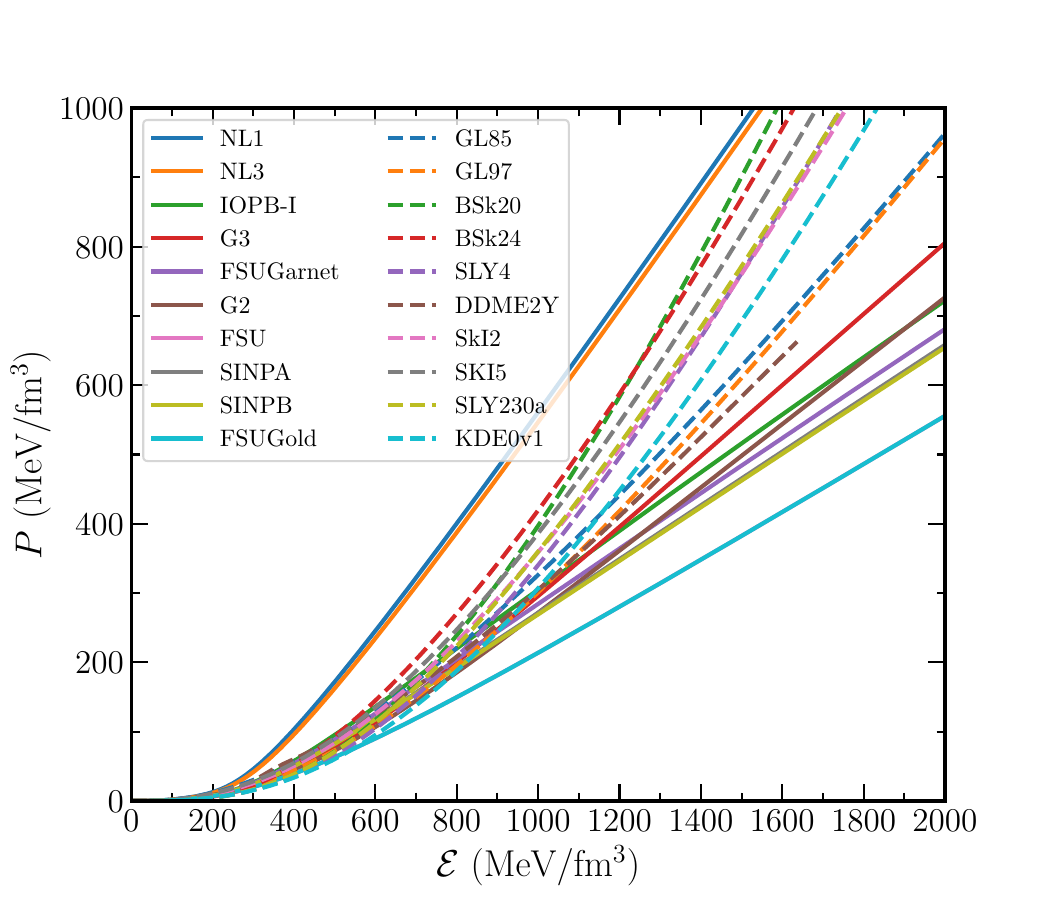}%%
\includegraphics[width=0.52\textwidth]{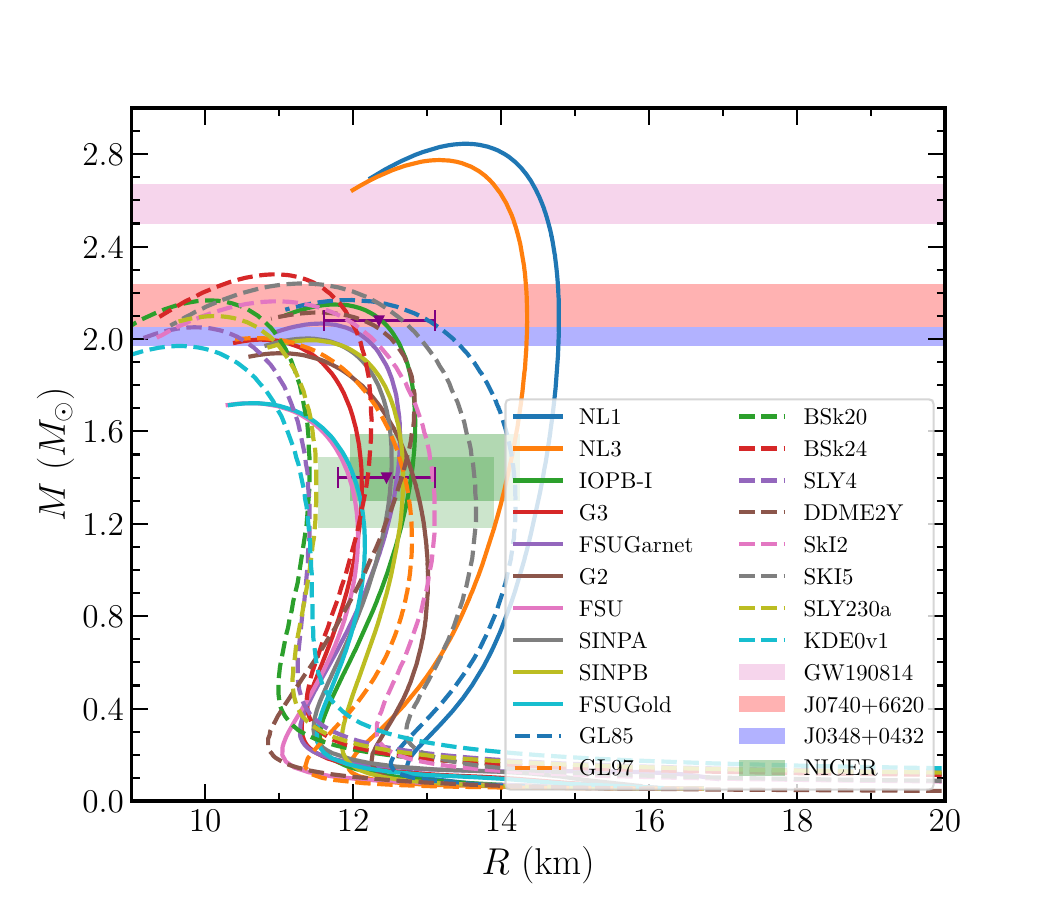}
\caption{{\it Left:} Some selected EOSs from RMF, SHF, and DD-RMF sets are shown to know their behavior mainly at the core part. {\it Right:} Mass and radius relations for selective parameter sets. The recent observational constraints from GW190814 and different pulsars J0740+6620 and J0348+0432 are shown with horizontal bars. The mass-radius relation observed by the NICER from two different analyses of pulsar PSR J0030+0451 is also shown \cite{Miller_2019, Riley_2019}. The magenta error bars represent the revised NICER constraints given by \cite{Miller_2021}.}
\label{fig_eos_mr}
\end{figure*}
%%%%%%%%%%%%

In Fig. \ref{fig_eos_mr} (left), we present a selection of EOSs from the RMF, SHF, and DD-RMF models. These models are chosen to demonstrate the EOS behavior, particularly within the core region of NSs. The figure depicts the pressure ($P$) as a function of energy density ($\mathcal{E}$) for a variety of parameter sets across different models. The lowest density segment of the EOS corresponds to the outer crust of the NS. The inner crust typically spans densities from approximately $3 \times 10^{-4}$ to $8 \times 10^{-2}$ fm$^{-3}$, while the core region commences around $2 \times 10^{-2}$ fm$^{-3}$ and extends to the NS center, where densities can reach 5–6 times the nuclear matter saturation density. At lower densities, all EOSs exhibit similar behavior owing to the adoption of a unified crust EOS (for both the outer and inner crust), specifically the BCPM EOS as described in Ref. \cite{BKS_2015}. The unified crust EOS describes the neutron star from outer crust to inner crust.  The outer crust does not affect the properties of the NS. Though the maximum mass of the NS and the corresponding radius are not affected by the inner crust, the canonical radius and tidal deformability  slightly vary  for different inner crust EOS \cite{RATHER21}. Here we have used the same unified crust EOS for all the parameter sets, so there is a little possibility that this correlation will be affected by the crust EOS. Secondly, this is the correlation between the properties, which mostly depends on the inner core EOS.  

Moreover, one can refer to unified EOSs as provided in Ref. \cite{Fortin_2016}. At higher densities, the EOSs diverge significantly due to differences in the underlying physics and parameter choices of each EOS model. This divergence reflects the extensive range of nuclear matter parameters covered by these models, facilitating the establishment of correlations with greater precision across different nuclear densities. The primary objective of utilizing a large number of parameter sets is to encompass a broad spectrum of nuclear matter parameter values, thereby offering comprehensive coverage for the study of NS properties. By analyzing this extensive array of EOSs, we aim to discern correlations between various quantities, thereby advancing our comprehension of the physics governing neutron stars.
%%%%%%%%%%%%%%%%%%%%%%%%%%%%%%%%%%%%%%%%
\subsection{Mass-Radius relations}
%%%%%%%%%%%%%%%%%%%%%%%%%%%%%%%%%%%%%%%%
In Fig. \ref{fig_eos_mr} (right), the mass-radius relations for NSs derived for selected EOSs from RMF, SHF, and DD-RMF models are presented. Recent observational constraints from gravitational wave events and various pulsar measurements are depicted as shaded regions and horizontal bars. Observational constraints from PSR J1614-2230 \cite{demo10}, PSR J0348+0432 \cite{anto13}, PSR J0740+6620 \cite{Cromartie_2019}, and the secondary component of the GW190814 event (with a mass range of 2.50--2.67 $M_{\odot}$) are superimposed for comparison. The GW190814 event has sparked considerable debate, with hypotheses ranging from it being a binary black hole merger \cite{Fattoyev_2020,das2020bigapple}, the heaviest known NS \cite{Tan_2020,Huang_2020}, an NS with dark matter admixture \cite{DasPRD_2021}, or even a super-fast pulsar \cite{Zhang_2020}. The NICER results, providing constraints on the mass-radius relation from the X-ray study of the millisecond pulsar PSR J0030+0451, are also encompassed. The results by Miller et al. \cite{Miller_2019} and Riley et al. \cite{Riley_2019} offer mass-radius constraints of $M=1.44_{-0.14}^{+0.15} \ M_{\odot}$ and $R=13.02_{-1.06}^{+1.24}$ km, and $M=1.34_{-0.16}^{+0.15} \ M_{\odot}$ and $R=12.71_{-1.19}^{+1.14}$ km, respectively, as indicated by the two shaded boxes. 

Among the RMF parameter sets, some, such as GL97, GL85, TM2, FSU2, IUFSU*, SINPA, SINPB, G1, G2, G3, IOPB-I, FSUGarnet, and FSU2R, match the range of observed maximum masses. Most RMF parameter sets meet the NICER observation constraints for the mass-radius relation, implying that the NICER data cannot exclude many EOSs based solely on mass-radius measurements. The NICER data pertain to the radius measurement of an NS with a mass around $1.4 \ M_\odot$, which does not correspond to a very high central density. The central density of a canonical NS generally lies within 2--3 times nuclear saturation density. Since all EOSs are calibrated to the saturation properties of nuclear matter, they exhibit convergent behavior up to a few times the saturation density. This is significant as the properties of smaller NSs, due to their lower central densities, show a strong correlation with the nuclear matter properties \cite{Carriere_2003}. Some parameter sets, such as HS, LA, LZ, NL-SH, NL1, and IUFSU, do not fit the NICER mass-radius constraints. Nevertheless, only eight parameter sets, including GM1, GL85, FSU2, IUFSU*, G1, IOPB-I, FSUGarnet, and FSU2R, conform to both the NICER and recent maximum NS mass constraints. For the SHF and DD-RMF models (right panel), the mass and radius of NSs are also illustrated in Fig. \ref{fig_eos_mr}. Except for DDLZ1, DDME1, DDME2, and DD2, all other parameter sets satisfy the maximum mass constraints imposed by different pulsar measurements. The NICER limits are fulfilled by all parameter sets except KDE0v1 and the SLY family. Notably, the maximum mass predicted by DDLZ1 falls within the constraints provided by the GW190814 event, indicating a potential EOS candidate for describing the secondary component of this event.

%%%%%%%%%%%%%%
\begin{figure*}
\centering
\includegraphics[width=0.51\textwidth]{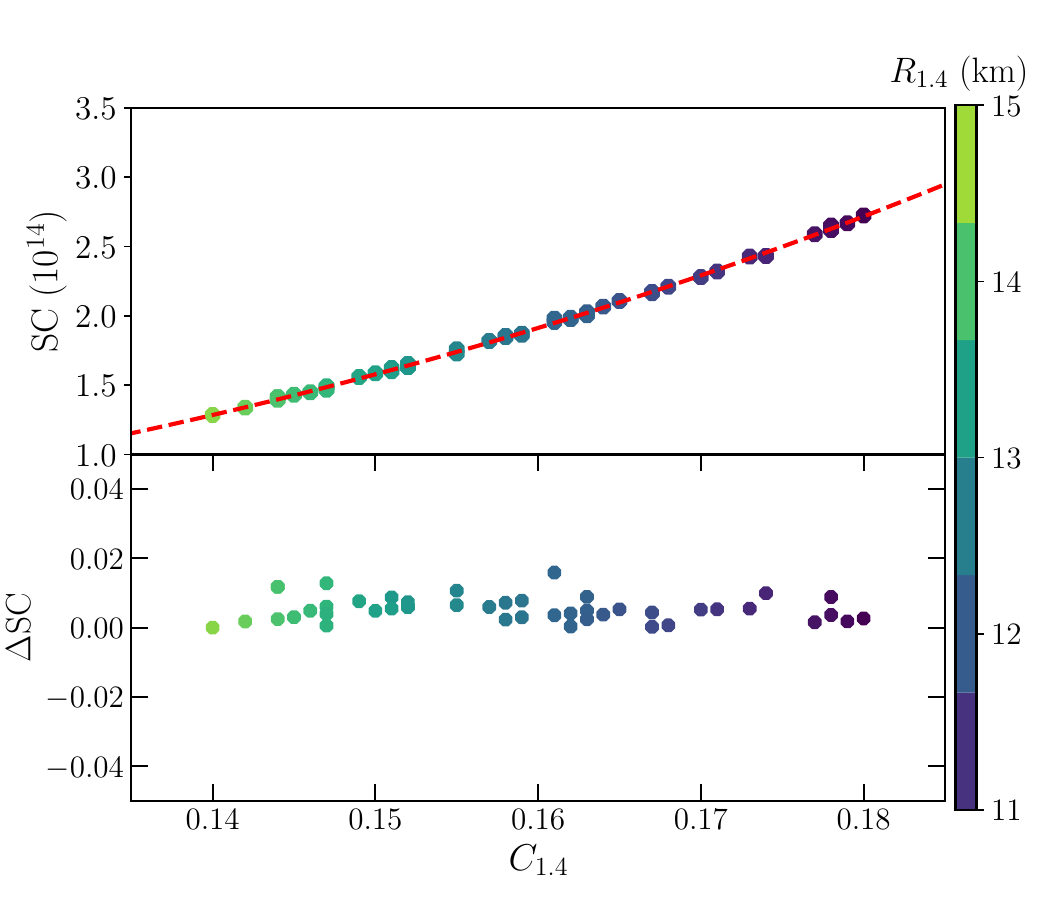}%
\includegraphics[width=0.51\textwidth]{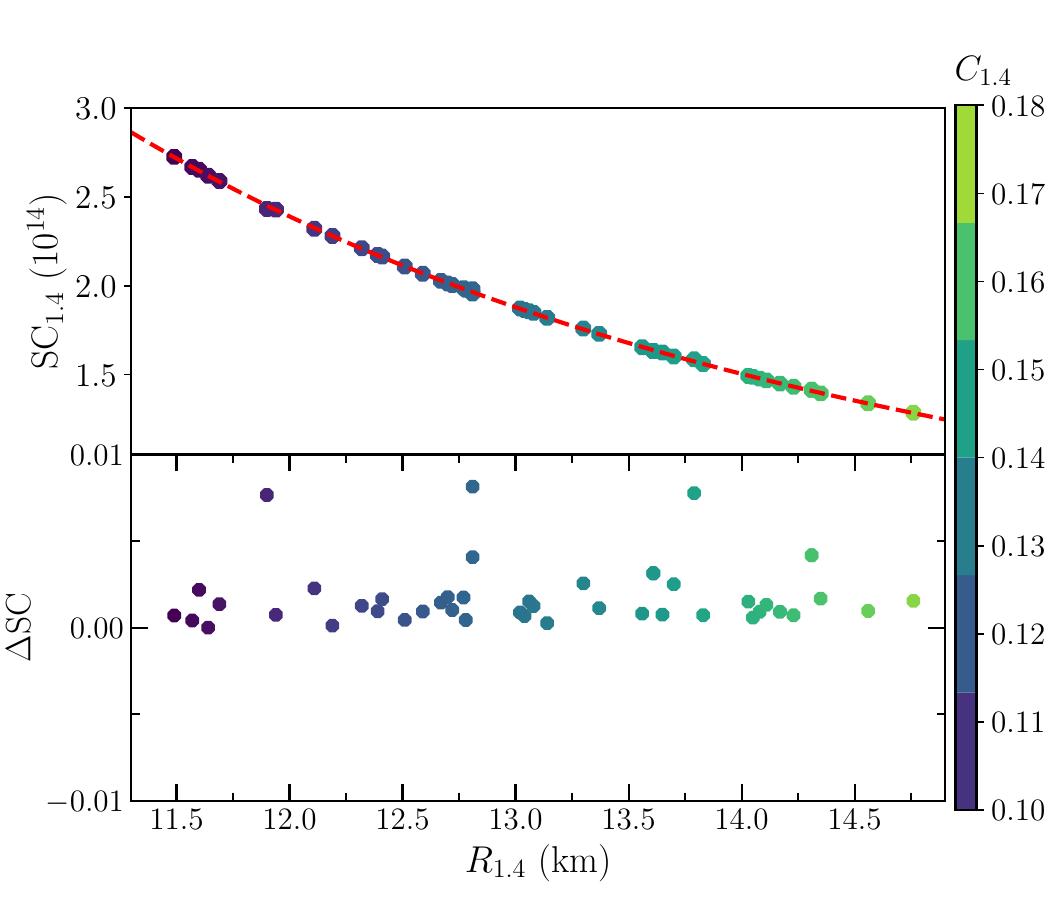}
\caption{{\it Left:} SC as a function of compactness for the canonical stars for various EOSs. {\it Right:} SC as a function of radius.}
\label{fig_curv_com_rad}
\end{figure*}
%%%%%%%%%%%%
\subsection{Relation between SC and Compactness ($C$)}

The compactness of an NS is defined as the ratio of its mass to its corresponding radius. In Fig. \ref{fig_curv_com_rad} (left), we plot the SC as a function of $C$ for canonical stars with different EOSs. The SC values are computed for a neutron star with a mass of 1.4 $M_{\odot}$ (denoted as ${\rm SC}_{1.4}$) and plotted against the compactness $C_{1.4}$. The data points are color-coded by the radius $R_{1.4}$ of the canonical star, as indicated by the color bar on the right side.

A least-squares fitting method is applied to find an approximate relationship between ${\rm SC}_{1.4}$ and $C_{1.4}$:

\begin{equation}
{\rm SC}_{1.4} = a \cdot C_{1.4}^b,
\end{equation}

where the fitting parameters are $a = 450.770$ and $b = 2.9806$. The dashed red line represents the best fit. The lower panel shows the residuals ($\Delta{\rm SC}$) for this fit, calculated using:

\begin{equation}
\Delta{\rm SC} = \frac{|{\rm SC} - {\rm SC}_{\rm fit}|}{{\rm SC}_{\rm fit}}.
\end{equation}

\begin{figure*}
\centering
\includegraphics[width=0.6\textwidth]{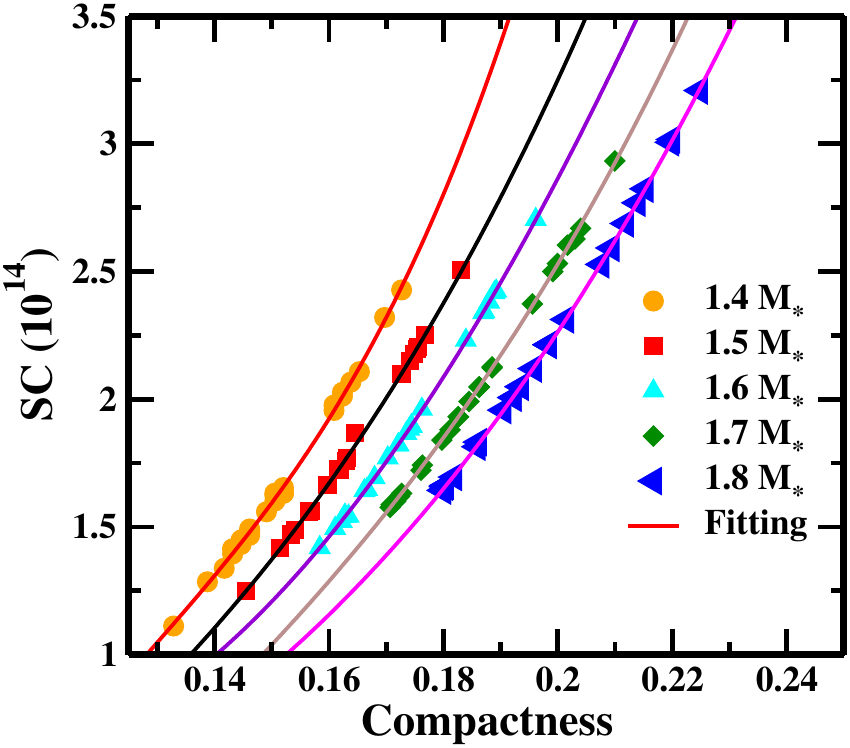}
\caption{The correlation between SC and compactness for considered parameter sets is
shown for different masses of the NS.}
\label{curv_com_mass}
\end{figure*}

The correlation coefficient ($r^2$) for this fit, calculated using the formula \cite{Chicco_2021}:

\begin{equation}
r^2 = 1 - \frac{\text{sum of squares of residuals}}{\text{total sum of squares}},
\end{equation}

is found to be 0.99, indicating a very strong correlation between $SC_{1.4}$ and $C_{1.4}$. The correlation of SC and compactness for different masses with the corresponding parameter set is also shown in Fig. \ref{curv_com_mass}, and shows that with an increase in mass, the correlation between SC and compactness gradually decreases. This correlation is not affected by the crust, as the crust contributes very little to the bulk properties of the neutron star.

\subsection{Relation between SC and Radius ($R$)}
Similarly, we investigate the relationship between SC and the radius of the NS for different EOSs, as shown in Fig. \ref{fig_curv_com_rad} (right). The SC values are again calculated for canonical stars with a mass of 1.4 $M_{\odot}$, denoted as $SC_{1.4}$, and plotted against the radius $R_{1.4}$. Data points are color-coded by compactness $C_{1.4}$, as shown by the color bar on the right side.

The data is fitted with the following equation:

\begin{equation}
{\rm SC}_{1.4} = \frac{c}{R_{1.4}^d},
\end{equation}

where the fitting coefficients are $c = 4179.953$ and $d = 3.00486$. The red dashed line shows the best fit and the lower panel shows the residuals, $\Delta{\rm SC}$. The correlation coefficient is found to be 0.99, which also indicates a strong correlation between $SC_{1.4}$ and $R_{1.4}$.

\subsection{Relation between SC and dimensionless tidal deformability ($\Lambda$)}

When an NS is in the gravitational field of its companion, it becomes deformed due to tidal forces. The tidal deformability of an NS is an important quantity that characterizes this deformation and is defined in the literature as the dimensionless tidal deformability, $\Lambda$, given by \cite{Hind08,hind10, KumarTide_2017, das2020bigapple}:

\begin{equation}
\Lambda = \frac{2}{3} k_2 C^5,
\end{equation}

where $k_2$ is the tidal Love number and $C$ is the compactness of the star.
% {\color{blue} $k_2$ is a dimensionless parameter that measures the amount of deformation caused due to external tidal field.  

The $k_2$ is defined as \cite{Hind08,hind10}
% \begin{equation}
%     k_2 = \frac{3}{2} G \lambda R^{-5},
% \end{equation}
% where $\lambda$ is a coefficient constant related to $l = 2$. The quadrupolar tidal love number $k_2$ for a spherically symmetric neutron star is given by 
\begin{eqnarray}
&&   k_2 = \frac{8C^5}{5} (1-2C^2)[2+2C(y-1)-y]\{ 2C[6-3y+3C(5y-8)]\nonumber\\
&&  + 4C^3 [13-11y+C(3y-2) + 2C^2 (1+y)] \nonumber\\
&&  + 3(1-2C^2) [2-y+2C(y-1)]log(1-2C)\}^{-1},
\end{eqnarray}

where $y$ can be obtained by solving the differential equation as given in Refs. \cite{Hind08,hind10}

The dimensionless tidal deformability has been widely used in gravitational wave studies to extract information about the internal structure of neutron stars from events such as GW170817 and GW190814.

%%%%%%%%%%%%%%
\begin{figure}
\centering
\includegraphics[width=0.7\textwidth]{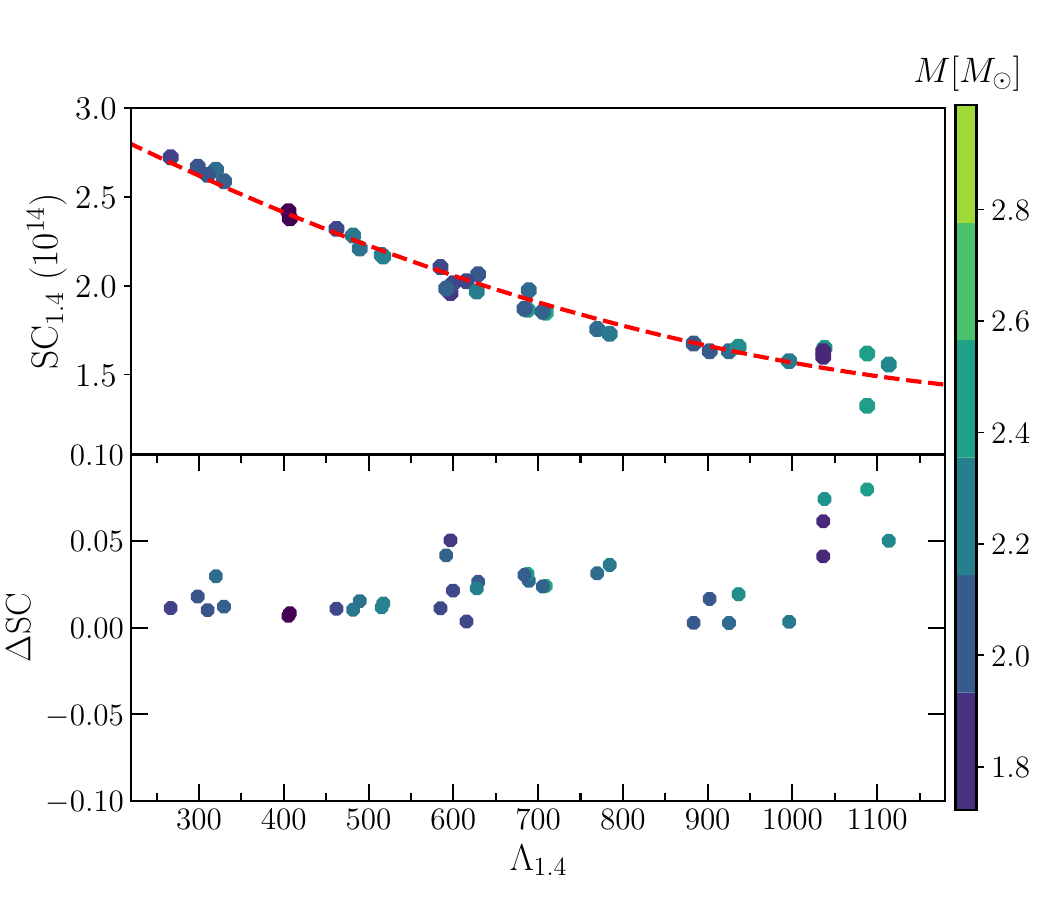}
\caption{The SC as the function of compactness for the canonical star for various EOSs.}
\label{fig_curv_tidal}
\end{figure}
%%%%%%%%%%%%
Figure \ref{fig_curv_tidal} shows the SC of canonical neutron stars with a mass of 1.4 $M_{\odot}$ (denoted as ${\rm SC}_{1.4}$) plotted as a function of the dimensionless tidal deformability ($\Lambda_{1.4}$) for various EOSs. The data points are color-coded according to the mass of the star, as indicated by the color bar on the right. To quantify the relationship between ${\rm SC}_{1.4}$ and $\Lambda_{1.4}$, we perform a polynomial fit using the following equation:

\begin{equation}
{\rm SC}_{1.4} = e_0 + e_1 \Lambda_{1.4} + e_2 \Lambda_{1.4}^2,
\end{equation}

where the coefficients are determined to be $e_0 = 3.35$, $e_1 = -2.71 \times 10^{-3}$, and $e_2 = 9.26 \times 10^{-7}$. The best fit is shown as a dashed red line in the upper panel of Fig. \ref{fig_curv_tidal}. The residuals, defined as

\begin{equation}
\Delta{\rm SC} = \frac{{\rm SC} - {\rm SC}_{\rm fit}}{{\rm SC}_{\rm fit}},
\end{equation}

are displayed in the lower panel of Fig. \ref{fig_curv_tidal}. The residuals are generally small, indicating a good fit with the proposed polynomial function. Using the tidal deformability constraints from the GW170817 event, where $\Lambda_{1.4} = 190_{-120}^{+390}$, we can predict the corresponding values of ${\rm SC}_{1.4}$. From this range, we obtain:

\begin{equation}
{\rm SC}_{1.4} (10^{14}) = 2.87^{+0.30}_{-0.78}.
\end{equation}

Similarly, for the GW190814 event in the neutron star-black hole scenario, the tidal deformability is constrained to $\Lambda_{1.4} = 616_{-158}^{+273}$ \cite{RAbbott_2020}. From this constraint, we derive the corresponding SC values as:

\begin{equation}
{\rm SC}_{1.4} (10^{14}) = 2.03^{+0.27}_{-0.36}.
\end{equation}

These results indicate that the GW190814 event provides a tighter constraint on the surface curvature of neutron stars compared to the GW170817 event. The strong correlation between SC and $\Lambda$ demonstrates that surface curvature can be a useful parameter for studying neutron star properties and provides valuable insight into the effects of tidal interactions on NS structure.

%%%%%%%%%%%%%%%%%%%%%
\section{Conclusions}
\label{conclusion}
%%%%%%%%%%%%%%%%%%%%%

In this study, we computed the mass, radius, SC, and compactness of NS using a variety of EOSs, specifically those based on RMF, SHF, and DDRMF models. We analyzed 26 RMF, 15 SHF, and 6 DDRMF EOSs to cover a broad spectrum of nuclear saturation properties. Most of these EOSs satisfy the constraints set by the maximum mass and NICER observations, with a few exceptions.

Our results show that the maximum mass and canonical radius predictions from the selected EOSs are consistent with constraints from NICER, GW190814, and the observed maximum masses of neutron stars. We identified a subset of EOSs that simultaneously obey the constraints from NICER, GW170817, and the recent maximum mass limits. However, our analysis also reveals that some EOSs, while capable of reproducing the maximum mass of a supermassive neutron star (which could potentially be the smallest black hole), fail to satisfy other constraints concurrently. Additionally, we explored the curvature of neutron stars using the RMF and SHF EOSs and discovered a significant correlation between the SC and the compactness of different neutron star masses. For the canonical 1.4 $M_{\odot}$ neutron star, this relationship follows a cubic correlation with a correlation coefficient of 0.99, indicating a very strong correlation. A similar correlation is observed between SC and the neutron star's radius, also with a correlation coefficient of 0.99 for the canonical star.

Furthermore, we derived a universal relation between the $\Lambda$ and SC of neutron stars. This universal relation allows us to constrain the SC based on observational data. For example, using the tidal deformability constraint from GW170817, $\Lambda_{1.4} = 190_{-120}^{+390}$, we estimate the surface curvature as SC$_{1.4} (10^{14}) = 2.87^{+0.30}_{-0.78}$. Similarly, the tidal deformability constraint from the secondary component in the GW190814 event in the NSBH scenario, $\Lambda_{1.4} = 616_{-158}^{+273}$ \cite{RAbbott_2020}, provides a tighter limit on SC, yielding SC$_{1.4} (10^{14}) = 2.03^{+0.27}_{-0.36}$. These findings suggest that the GW190814 event imposes stricter constraints on the SC compared to GW170817.

%%%%%%%%%%%%%%%%%%%%%%
%\begin{acknowledgments}
\section*{Acknowledgements}
S. K. Biswal acknowledges the support from the Science and Engineering Research Board,
Department of Science and Technology, Government of India, with grant no. CRG/2022/005378. B.K. acknowledges partial support from the Department of Science and Technology, Government of India, with grant no. CRG/2021/000101.
%\end{acknowledgments}
%%%%%%%%%%%%%%%%%%%%%%%%%%
\bibliography{correlation}
\bibliographystyle{ws-ijmpe}
%%%%%%%%%%%%%%%%%%%%%%%%%%
\end{document}